# Spin wave spectrum of a magnonic crystal with an isolated defect

V.V. Kruglyak

School of Physics, University of Exeter, Stocker Road, Exeter, EX4 4QL, UK

M.L. Sokolovskii, V.S. Tkachenko, and A.N. Kuchko

Donetsk National University, 24, Universitetskaya Street, Donetsk, 83055, Ukraine

**Abstract**

Real magnonic crystals – periodic magnetic media for spin wave (magnon) propagation – may contain some defects. We report theoretical spin wave spectra of a one dimensional magnonic crystal with an isolated defect. The latter is modeled by insertion of an additional layer with thickness and magnetic anisotropy values different from those of the magnonic crystal constituent layers. The defect layer leads to appearance of several localized defect modes within the magnonic band gaps. The frequency and the number of the defect modes may be controlled by varying parameters of the constituent layers of the magnonic crystal.

The translational symmetry (spatial periodicity) is one of the most important notions in the modern understanding of nature. It determines the conservation of momentum of a material body in the free space and of the quasi-momentum of electrons in crystals. Most of electronic, magnetic and optical properties of solids are directly determined by their band structure, which directly results from the electron quasi-momentum conservation. The spectrum of electrons in solids splits into so-called bands – energy regions in which electron propagation is allowed. There exist also band-gaps – energy regions in which there are no available electronic states, and so the electron propagation is prohibited. Structures with artificial translational symmetry have been created to design objects with properties that otherwise do not exist in nature. The examples of these include photonic crystals[1], semiconductor superlattices[2], plasmonic[3] and phononic crystals[4]. The band spectrum appears to be important even for such exotic objects as carbon nanotubes in a transverse electric field[5].

Magnetic materials with periodically modulated properties (magnetic superlattices) are known to possess such unique properties as giant magnetoresistance (GMR)[6], large out-of-plane magnetic anisotropy[7], resonant absorption of microwaves[8], and magnetic field controlled photonic band gaps[9]. These materials have also been used as retardation lines in which magnetostatic waves[10] are used as carriers of signal[11]. Such periodic magnetic structures considered as a medium of magnon (spin wave) propagation have been called magnonic crystals. Similarly to the above mentioned artificial crystals, the spectrum of a magnonic crystal is strongly influenced by the presence of magnonic band gaps in which magnon propagation is forbidden[12].

In real magnonic crystals, the presence of defects can lead to a local modification of the values of such parameters as magnetic anisotropy, exchange stiffness, saturation magnetization, and hence can break the translational symmetry of the magnonic crystal and affect its spectrum. Nikitov et al showed that a local modification of the thickness of one of the layers of a magnonic crystal leads to occurrence of a discrete level within the band gap[13]. In more detail, the defect modes have been studied for photonic and phononic crystals[14-16]. In this work, we investigate the spectrum of a magnonic crystal that contains an isolated defect of the magnetic anisotropy value. The graphical method developed in Ref. 12 is modified for analysis and control of the defect modes within an arbitrary band gap as well as in the spectral region below the spin wave activation frequency.

Let us consider an infinite magnonic crystal that consists of periodically repeated thin film layers of two types, i.e. …ABABABAB…, as shown in Figure 1. The layers differ by their thicknesses ($a$ and $b$) and values of the uniaxial anisotropy ($\beta_a$ and $\beta_b$). The values of the spontaneous magnetization $M_S$ and the exchange parameter $\alpha$ are assumed to be constant throughout the material. It is also assumed that the easy axis of the uniaxial anisotropy, the internal magnetic field $\mathbf{H}_i$ [10] and the static magnetization direction are all perpendicular to layers of the magnonic crystal. The defect is modeled as a layer with thickness $d$ and an anisotropy value $\beta_d$ inserted into the magnonic crystal. The Cartesian coordinate system is defined so that its Z axis is parallel to the easy axis.

The dynamics of magnetization $\mathbf{M}(\mathbf{r},t)$ is described by the Landau-Lifshitz equation[10]

$$\frac{\partial \mathbf{M}}{\partial t} = -g[\mathbf{M} \times \mathbf{H}_E], \qquad (1)$$

where $g$ is the gyromagnetic ratio ($g>0$) and $\mathbf{H}_E$ is the effective magnetic field. In the following, we will restrict ourselves to discussion of spin waves that propagate parallel to the static magnetization. Then for the chosen geometry, dynamic magneto-static fields do not contribute to the effective field[10], and we can write for the latter

$$\mathbf{H}_E = [H + \beta(\mathbf{Mn})]\mathbf{n} + \alpha \Delta \mathbf{M}, \qquad (2)$$

where $H$ is the magnitude of the internal magnetic field[10] and **n** is the unit vector parallel to the Z axis.

Let us consider small deviations **m** (**r**,$t$) of the magnetization from the ground state, i.e. a uniform magnetization parallel to the easy axis. For this purpose, we represent magnetization as

$$\mathbf{M}(\mathbf{r},t) = M_S \mathbf{n} + \mathbf{m}(\mathbf{r},t), \text{ where } |\mathbf{m}| \ll M_S. \tag{3}$$

Introducing notations $m_{\pm} = m_x \pm i m_y$ and seeking solutions in the form of harmonic waves $m_{\pm}(\mathbf{r},t) = m(z)\exp\{\pm i\omega t\}$, we obtain the following linearized equation for $m(z)$

$$\frac{d^2 m(z)}{dz^2} + \left(\frac{\Omega - h - \beta(z)}{\alpha}\right) m(z) = 0, \tag{4}$$

where $\Omega = \omega/gM_S$ and $h = H/M_S$ are the dimensionless frequency and magnetic field. Within each of the uniform layers of the magnonic crystal, equation (4) admits solutions

$$m(z) = A_\nu^{(+)} \exp\{+ik_\nu z\} + A_\nu^{(-)} \exp\{-ik_\nu z\}, \tag{5}$$

where index $\nu$ denotes different layer types, being a, b, or d. $A_\nu^{(\pm)}$ are the spin wave amplitudes in the layers. The wave number of the spin wave within a particular layer is given by

$$k_\nu = \sqrt{\frac{\Omega - h - \beta_\nu}{\alpha}}. \tag{6}$$

At interfaces, functions (5) must satisfy the exchange boundary conditions[17] requiring the continuity of the magnetization $m(z)$ and its derivative $dm(z)/dz$.

As formulated above, the problem of finding the magnonic spectrum and the defect mode frequencies within the spectral band gaps is similar to that considered by

Tamura in Ref. 15 for the case of a phononic crystal. Following the "transfer matrix" method described in detail in Ref. 15, we obtain for the spin wave spectrum of a magnonic crystal without defects

$$2\cos(\kappa l) = F = \lambda + \mu, \quad (7)$$

where $\kappa$ is the Bloch wave number, $l$ is the period of the magnonic crystal, and

$$\lambda = \cos(k_a a)\cos(k_b b) - \left(\frac{k_a}{k_b}\right)\sin(k_a a)\sin(k_b b),$$

$$\mu = \cos(k_a a)\cos(k_b b) - \left(\frac{k_b}{k_a}\right)\sin(k_a a)\sin(k_b b). \quad (8)$$

The defect mode frequencies are given by solutions of equation

$$\Delta(\Omega) = 2\gamma_0 \lambda_0 - \gamma_1 (\lambda_0 F - \sigma \zeta_0 - \sigma_0 \zeta) = 0, \quad (9)$$

where

$$\sigma = \frac{1}{k_a}\sin(k_a a)\cos(k_b b) + \frac{1}{k_b}\cos(k_a a)\sin(k_b b),$$

$$\zeta = -k_a \sin(k_a a)\cos(k_b b) - k_b \cos(k_a a)\sin(k_b b)$$

$$\lambda_0 = \cos(k_d d), \quad \sigma_0 = \frac{1}{k_d}\sin(k_d d), \quad \zeta_0 = -k_d \sin(k_d d), \quad (10)$$

$$\gamma_0 = -\frac{1}{\sqrt{F^2 - 4}}, \quad \gamma_1 = \gamma_0 \left[\frac{F - \sqrt{F^2 - 4}}{2}\right].$$

Equation (7) describes a spectrum with band gaps, such as shown in Figure 2 (a). The Brillouin zone boundaries are defined by condition $|F| = 2$, and are independent of the parameters of the defect. As shown in Figure 3, this facilitates mapping of the allowed bands (black) and the band gaps (white) on the $(\alpha k_a^2, \alpha k_b^2)$ plane in a manner similar to Ref. 12. Using coordinates $(\alpha k_a^2, \alpha k_b^2)$ instead of $(k_a, k_b)$ helps to consider the frequency region where either $k_a$, or $k_b$, or both, is purely imaginary. Also, in these coordinates, the "lines of spectra" $\Omega(\beta_a, k_a) = \Omega(\beta_b, k_b)$ [12] are straight lines at 45 degrees to the axes, and the width of the band gaps is simply equal to the distance between the points of intersection of the lines of spectra with the Brillouin zone boundaries (the boundaries between the black and white).

Within the band gaps, $|F| > 2$, and hence $\kappa$ is imaginary. In this case, the values of the parameters defined in (8) and (10) are all real. The frequency dependence of function $\Delta(\Omega)$ is shown in Figure 2 (b). The solutions of equation (9) define the frequency values of spin wave modes localized on the defect. While equation (9) is easily solved numerically, this does not allow one to predict how the spectrum, such as that shown in Figure 2 (a), changes when some or all of the parameters of the magnonic crystal are varied. The diagram in Figure 3 can again be used to circumvent this problem. Let us note that equation (9) does not contain the frequency explicitly, which allows us to draw the "defect lines", in which $\Delta(\Omega) = 0$, on the same diagram in Figure 3. Again, the points of intersection of the lines of spectra with the defect lines within the band gaps correspond to the defect modes. For example, one can clearly identify the four defect modes within the first band gap shown in Figure 2. Different depths of modulation of the anisotropy parameter will result in different lines of spectra. For example, if the depth of modulation of the anisotropy parameter is decreased, the line of spectra will shift towards the diagonal of the diagram in Figure 3. It is easy to see that this will result in a decrease in the size of the band gaps and the number of defect modes within them.

Technically, magnonic crystals with nearly constant saturation magnetization and exchange parameter values but the anisotropy constant modulated could be made of Co-P alloy[18,19]. In the more general case of the modulation of several magnetic parameters, this graphical technique can be used for investigation of the associated effects in the manner described in Refs. 12,17. We also expect this technique to be applicable for studying effects caused by the presence of the interface anisotropy between the defect and the superlattice[20], by the defect-superlattice symmetry/asymmetry (ABDAB as opposed to ABDBA), by the presence of the spin wave damping[21,22], and many other effects, which are however beyond the scope of the present paper. Finally, we note that our method could perhaps be applied to fields of physics other than magnetism, in particular, to those discussed in Refs. 1-5.

In summary, we have developed a graphical technique by which to study defect spin wave (magnon) modes within imperfect magnonic crystals. The technique may be especially useful in design of magnonic crystals for use in spin wave magnetic logic devices, such as those proposed in Ref. 23, in which a defect is created artificially to induce a phase shift to propagating spin waves.

List of figure captions.

Figure 1. Schematic of a one-dimensional magnonic crystal with a defect is shown.

Figure 2 (a) SW spectrum of a one-dimensional magnonic crystal with a defect is shown for $H=0$, $\beta_a=0.5$, $\beta_b=1.5$, $\beta_d=0.8$, $a/\sqrt{\alpha}=b/\sqrt{\alpha}=2.5$, $d/\sqrt{\alpha}=40$. The horizontal dashed lines represents the boundaries of the first band gap. $\Omega_n^{-/+}$ are the boundaries of the allowed bands 1 and 2. 3 are the defect modes. (b) Function $\Delta(\Omega)$ is plotted for the same parameter values. The points of intersection of $\Delta(\Omega)$ with 0 determines the frequencies of the defect modes in (a).

Figure 3 The diagram for determination of the defect mode positions within the band-gaps in the spectrum of a magnonic crystal is presented with the band gaps shown by white and the allowed bands shown by black. The "defect lines" are shown by thin solid lines, and the thick line at 45 degrees is one of the "lines of spectra". The defect lines and the line of spectrum are plotted for the parameters values from Figure 2.

1	J.D. Joannopulos, R.D. Meade, and J.N. Win, *Photonic Crystals* (Princeton Univ. Press, Princeton, 1995).

2	H.T. Grahn (Editor), *Semiconductor Superlattices: Growth & Electronic Properties* (World Scientific, Singapore, 1995).

3	W.L. Barnes, A. Dereux, and T.W. Ebbesen, Nature (London) **424**, 824 (2003).

4	M.S. Kushwaha, P. Halevi, L. Dobrzynski, and B. Djafari-Rouhani, Phys. Rev. Lett. **71**, 2022 (1993).

5	O.V. Kibis, D.G.W. Parfitt, and M.E. Portnoi, Phys. Rev. B **71**, 035411 (2005).

6	M.N. Baibich, J.M. Broto, A. Fert, F.N. Vandau, F. Petroff, P. Eitenne, G. Creuzet, A. Friederich, and J. Chazelas, Phys. Rev. Lett. **61**, 2472 (1988).

7	P.F. Carcia, A.D. Meinhaldt, and A. Suna, Appl. Phys. Lett. **47**, 178 (1985).

8	A. Saib, D. Vanhoenacker-Janvier, I. Huynen, A. Encinas, L. Piraux, E. Ferain, and R. Legras, Appl. Phys. Lett. **83**, 2378 (2003).

9	I.L. Lyubchanskii, N.N. Dadoenkova, M.I. Lyubchanskii, E.A. Shapovalov, and T.H. Rasing, J. Phys. D: Appl. Phys. **36**, R277 (2003).

10	A.G. Gurevich and G.A. Melkov, *Magnetization Oscillations and Waves* (Chemical Rubber Corp., New York, 1996).

11	Y.V. Gulyaev, S.A. Nikitov, L.V. Zhivotovskiĭ, A.A. Klimov, P. Tailhades, L. Presmanes, C. Bonnibgue, C.S. Tsai, S.L. Vysotskiĭ, and Y.A. Filimonov, JETP Lett. **77**, 567 (2003).

12	V.V. Kruglyak and A.N. Kuchko, Physica B **339**, 130 (2003).

13	S.A. Nikitov, Ph. Tailhades, and C.S. Tsai, J. Magn. Magn. Mater. **236**, 320 (2001).

14	S.Y. Vetrov and A.V. Shabanov, JETP **93**, 977 (2001).

15	S.I. Tamura, Phys. Rev. B **39**, 1261 (1989).

16	S. Mizuno and S. Tamura, Phys. Rev. B **45**, 13423 (1992).

17	V.V. Kruglyak, A.N. Kuchko, and V.I. Finokhin, Phys. Solid State **46**, 867 (2004).

18      R.S. Iskhakov et al, Fiz. Met. Metalloved. **56**, 85 (1983).

19      R.S. Iskhakov et al, Tech. Phys. Lett. **29**, 263 (2003).

20      K.Y. Guslienko, Fiz. Tverd. Tela **37**, 1603 (1995).

21      V.V. Kruglyak and A.N. Kuchko, J. Magn. Magn. Mater. **272-276**, 302 (2004).

22      V.V. Kruglyak, A.N. Kuchko, and V.Y. Gorobets, Metallofiz. Noveish. Tekhn. **26**, 579 (2004).

23      B. Leven et al, 50[th] MMM Conference (2005), Talk FF-01 (unpublished).

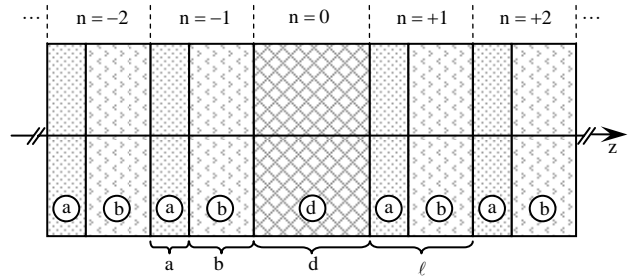
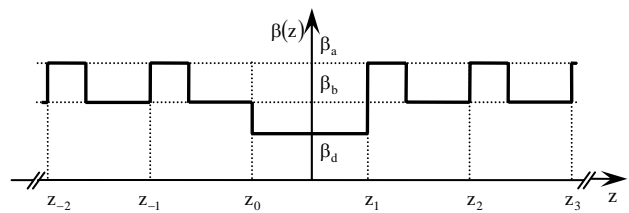

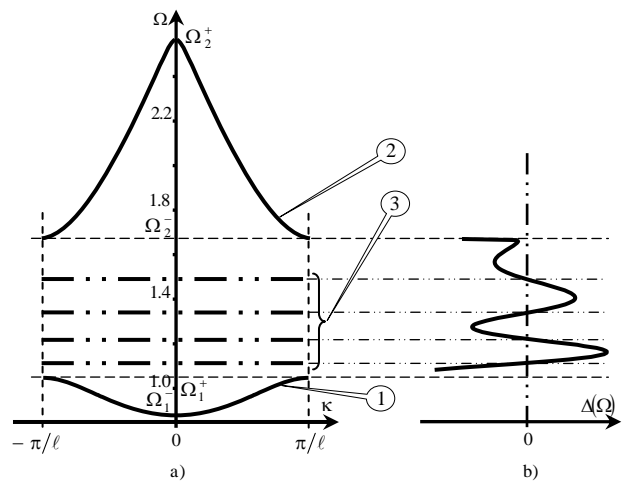

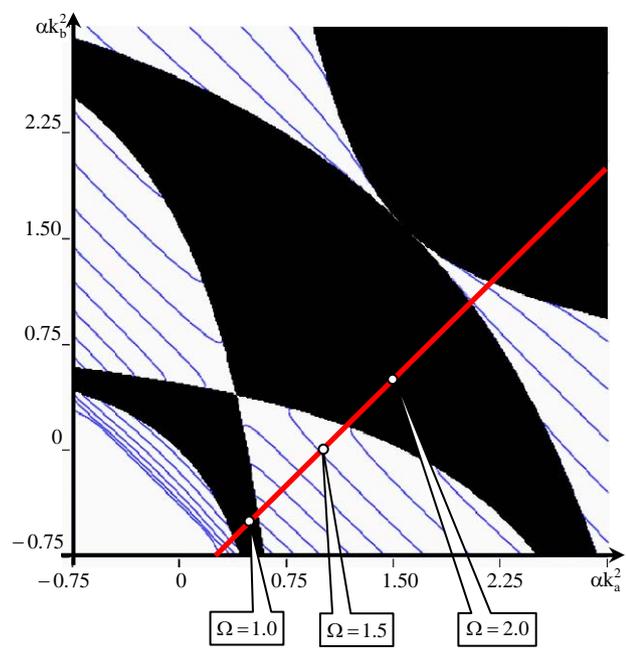